\documentstyle[11pt,cs10,epsf]{article}

\begin{document}

\title{Beryllium, Lithium and Oxygen Abundances in F-type Stars}

\author{R. J. Garc\'\i a L\'opez\altaffilmark{1,2}, M. C. Dom\'\i nguez
Herrera\altaffilmark{1,2}, M. R. P\'erez de Taoro\altaffilmark{3}, C.
Casares\altaffilmark{4}, J. L. Rasilla\altaffilmark{1}, R.
Rebolo\altaffilmark{1}, and C. Allende Prieto\altaffilmark{1}}

\altaffiltext{1}{Instituto de Astrof\'\i sica de Canarias} 
\altaffiltext{2}{Departamento de Astrof\'\i sica, Universidad de La Laguna} 
\altaffiltext{3}{Museo de la Ciencia y el Cosmos de Tenerife}
\altaffiltext{4}{Departamento de F\'\i sica Te\'orica II, Universidad 
 Complutense de Madrid}

\setcounter{footnote}{4}

\begin{abstract}
Beryllium and oxygen abundances have been derived in a sample of F-type field
stars for which lithium abundances had been measured previously, with the aim
of obtaining observational constraints to discriminate between the different
mixing mechanisms proposed. Mixing associated with the transport of angular
momentum in the stellar interior and internal gravity waves within the
framework of rotating evolutionary models, appear to be promising ways to
explain the observations.
\end{abstract}


\section{Introduction}

We present new beryllium and oxygen abundances in old disk-population 
F-type field stars with available lithium measurements. These objects,
which are slightly evolved off the main sequence, are mainly located in the
effective temperature range for which a strong lithium depletion, the so
called ``Li gap'', is observed among cluster and field stars older than $\sim
10^8$ yr (see review by Proffitt \& Michaud 1991). 
\index{Li gap}

Li and Be are fragile elements which are easily destroyed by ({\it
p,$\alpha$}) nuclear reactions when the temperature in the stellar interior
reaches $\sim 2.5\times 10^6$ and $\sim 3.5\times 10^6$ K, respectively;
while oxygen, with a stronger nucleus, could be affected by transport
mechanisms associated with microscopic diffusion (Michaud 1988). Surface
abundances of these three elements, showing different levels of fragility,
are used to help to discriminate between different mixing mechanisms
proposed.
\index{Microscopic Diffusion}

\section{Stars Selected}

We set about measuring the Be and O abundances in a group of F-type stars with
available Li abundances derived by Balachandran (1990). Stellar parameters
($T_{\rm eff}$, $\log g$, [Fe/H], $M_V$, and $v\sin i$) were also taken from
this work. Their masses were estimated by dividing the stars into three similar
groups of metallicity and comparing their ($M_V$,$T_{\rm eff}$) values with
the evolutionary tracks of VandenBerg (1985) corresponding to
[Fe/H]$=0.00,\,-0.23$, and $-0.46$, respectively. 

The 51 stars observed span from 5900 to 7000 K in $T_{\rm eff}$, from $-0.65$
to 0.20 in [Fe/H], and from 0.9 to 1.7 $M_\odot$. Recently revised ages, based
on $M_V$ values derived from {\it Hipparcos} parallaxes (Ng \& Bertelli 1997)
are available for 17 stars in the sample (covering the whole range of
metallicity), showing that the ages of the stars are within the interval of 1.8
to 6.6 Gyr.

\section{Observations and Analysis}

A detailed abundance analysis via LTE spectral synthesis has been carried out
to derive beryllium abundances, using the \ion{$^9$Be}{2} $\lambda$ 3131
\AA\ doublet, in 21 stars observed with $\lambda /\Delta\lambda\sim 5\times
10^4$ and $\sim 4\times 10^4$, using the Utrecht Echelle Spectrograph at the 
4.2 m William Herschel Telescope, and the IACUB echelle spectrograph (McKeith
et al. 1993) at the 2.5 m Nordic Optical Telescope (NOT), respectively, of 
the Roque de los Muchachos Observatory (La Palma, Spain). Similar analyses
have been carried out previously for the Sun, metal-poor stars, and the Hyades
(Garc\'\i a L\'opez et al. 1995a,b), showing the reliability of the method.
The left panel of Figure 1 shows the observed and synthetic spectra for one
of the stars.
\index{Observatories!Roque de los Muchachos}
\index{*Hyades}
\index{*HR6541}
\begin{figure}[htb]
\plotfiddle{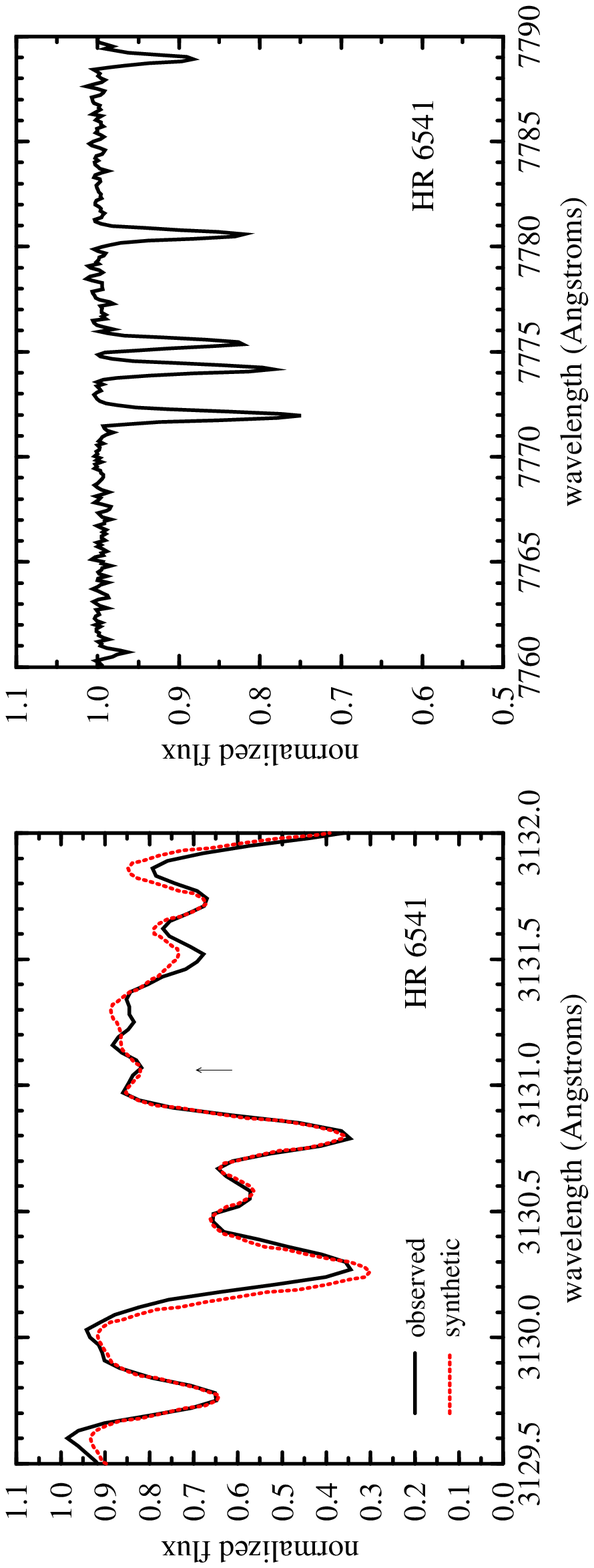}{5cm}{270}{54}{58}{-212}{340}
\caption[]{Left panel: fitting of a synthetic spectrum to the observed one in
the beryllium region of the star HR6541. The feature corresponding to the
\ion{Be}{2} $\lambda$ 3131.065 \AA\ line (weaker than the other Be line in
the doublet but more isolated) is indicated. Right panel: observed spectrum of
the same star corresponding to the \ion{O}{1} infrared triplet at
$\lambda\lambda$ 7771--7775 \AA.}
\label{fig1}
\end{figure}

Very recently, Balachandran \& Bell (1997) have argued that the continuous
opacity in the UV region could have not been fully accounted for in previous
works, providing abundances smaller than the actual ones. If this were the
case, there would be an overall change of scale in the beryllium abundances
derived from the \ion{Be}{2} $\lambda$ 3131.065 \AA\ line.
\index{*Sun}

Oxygen abundances of 37 stars have been derived using an NLTE analysis of the
\ion{O}{1} infrared triplet at $\lambda\lambda$ 7771--7775 \AA, following the
prescriptions employed by Garc\'\i a L\'opez et al. (1993). The observations
were carried out in two runs using IACUB at the NOT, with resolutions
$\lambda /\Delta\lambda\sim 2.9\times 10^4$ and $\sim 1.8\times 10^4$,
respectively. An example of the spectra in the \ion{O}{1} region is shown
in the right panel of Figure 1.   

\section{Abundances vs. Stellar Parameters}

Figure 2 shows the Be and Li abundances (where log (X)$=\log ({\rm
X/H})+12$), as well as the O abundances with respect to the Sun ([O/H]),
against stellar mass. All plots have been made with the same range of
abundances, and a large scatter can be seen within the Be and Li measurements
(filled circles) and upper limits (inverted open triangles), while the
dispersion in oxygen abundances is much smaller.
\begin{figure}[htb]
\plotfiddle{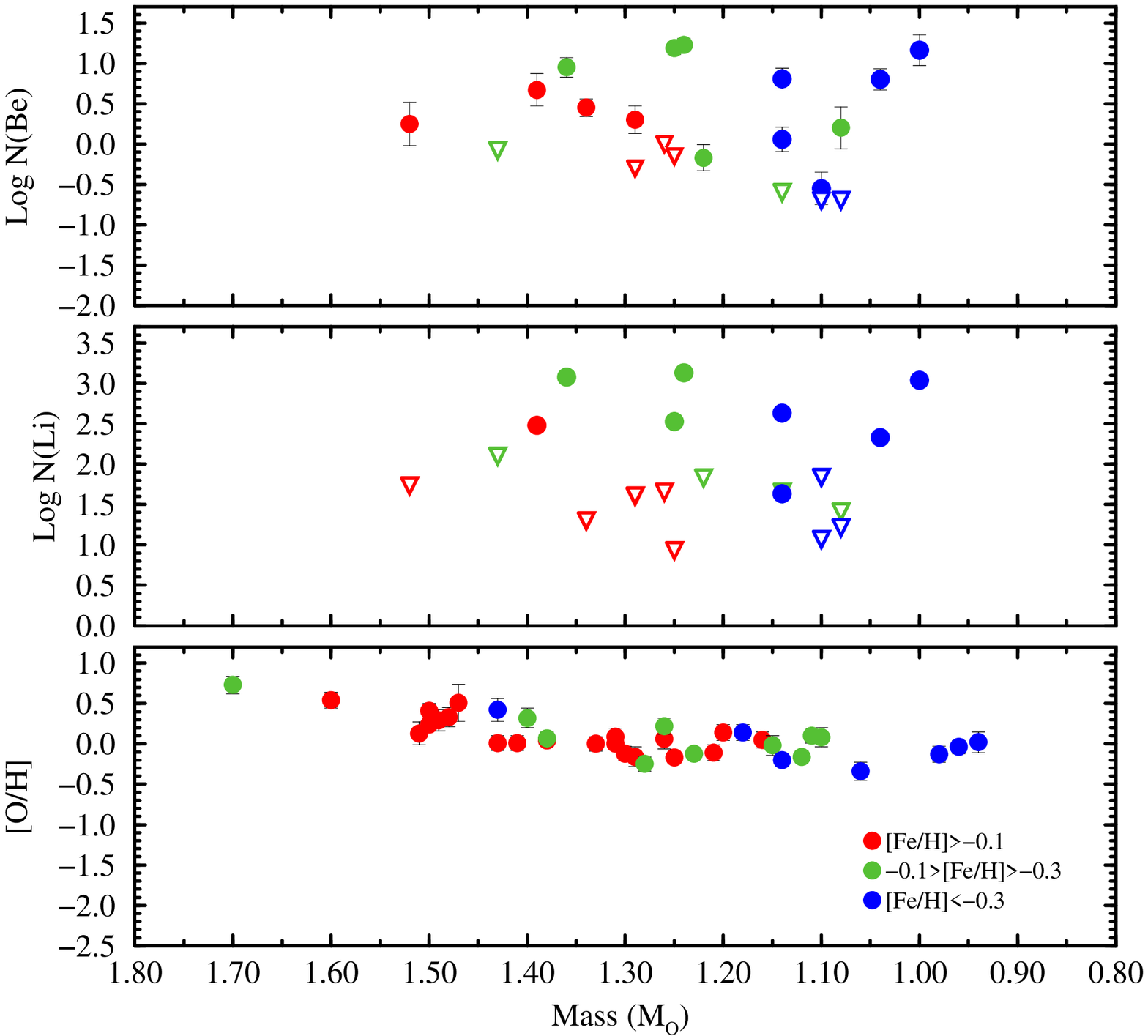}{10cm}{0}{54}{58}{-180}{0}
\caption{Beryllium, lithium and oxygen abundances (in the latter case with
respect to the Sun) against stellar mass for the F-type stars studied. Filled
circles represent detections and open inverted triangles upper limits. 
Different colors correspond to the three bins of metallicity in which the
sample has been divided. Note the large scatter for the Li and Be,
while O abundances show a tight correlation with mass.}
\label{fig2}
\end{figure}

All Be-depleted stars are also Li-depleted, as is expected from their
different nuclear reaction temperatures. HR6467 is the star with the least
Be detected (log N(Be)$=-0.55$), and there are eight stars with detections
in both Be and Li whose abundances are plotted in Figure 3. 
\index{*HR6467}
\begin{figure}[htb]
\plotfiddle{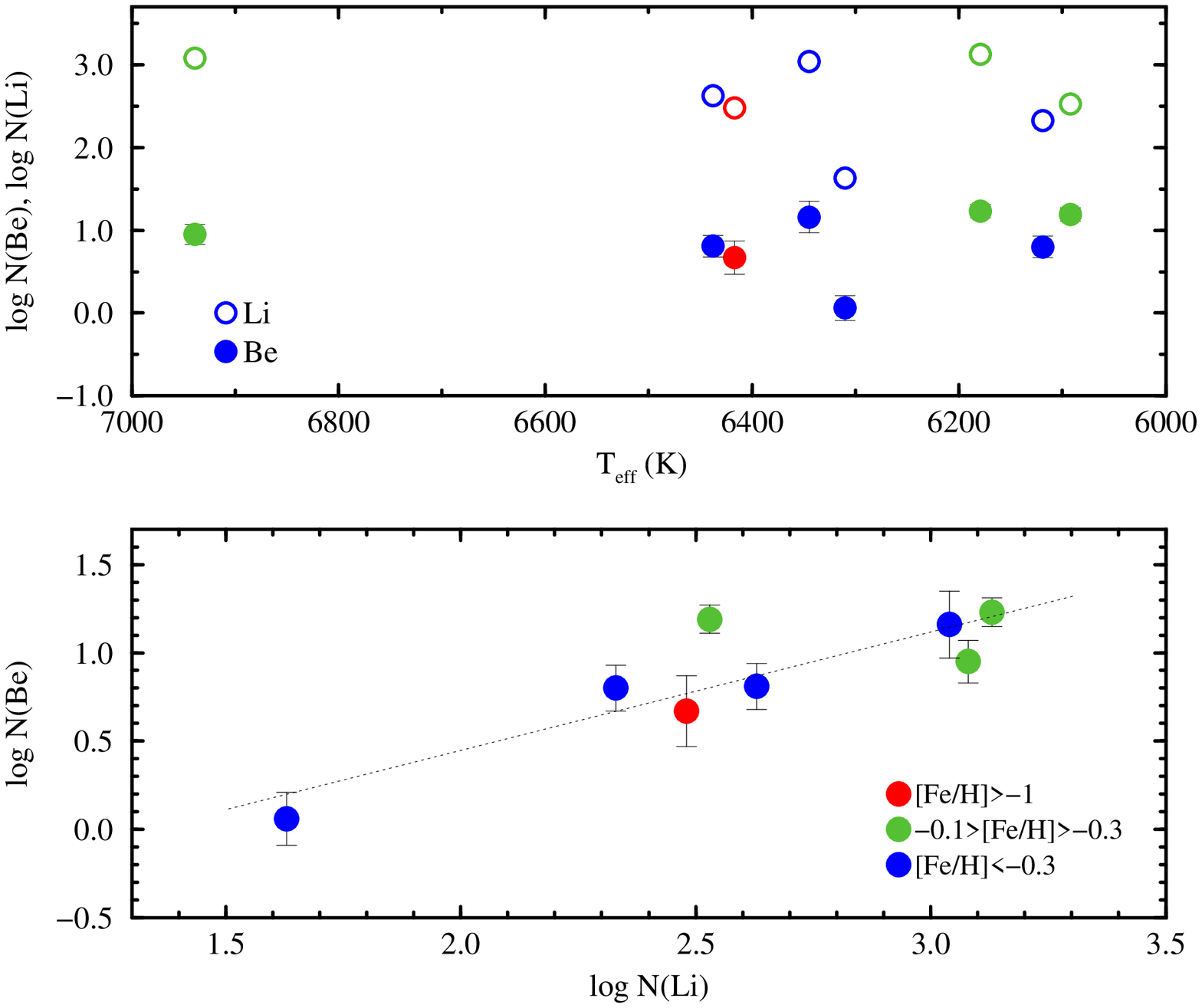}{10cm}{0}{54}{58}{-180}{0}
\caption{Upper panel: lithium and beryllium abundances against effective
temperature for those stars with clear detections of both elements. Lower
panel: beryllium against lithium abundances for these stars.}
\label{fig3}
\end{figure}
While the initial or ``cosmic'' lithium abundance of these F-type stars can
be estimated around log N(Li)$\sim 3.1-3.3$, the initial Be abundance for the
eight stars could vary, in principle, depending on their metallicity (due to
the steep relation observed between Be and [Fe/H] and associated with
spallation reactions between cosmic rays and C, N, and O nuclei in the
interstellar medium; e.g. Rebolo et al. 1995; Molaro et al. 1997). However,
given that their range in [Fe/H] is very small ($-0.30$ to 0.20) it is
conceivable that their initial abundances were not very different. If so, the
correlation seen in the lower panel of Figure 3 and the mean ratio log
(Li/Be)$=1.7\pm 0.3$ obtained from these stars could be used to constrain the
mechanism responsible for the Li and Be depletions among them. This ratio
would change to a smaller value if the UV continuous opacity were wrong as
suggested by Balachandran \& Bell (1997).

While there is no qualitative change in the distribution of Be or Li 
abundances against $T_{\rm eff}$ and mass, indicating that the depletion
mechanism is not related only to the stellar structure, the scatter of the
oxygen abundances decreases and a significant correlation appears when
plotting them against mass instead of against effective temperature. We have
no clear explanation for this at present, and the radiative accelerations on
oxygen computed by Gonz\'alez et al. (1995) for A- and F-type stars do not
even reproduce the observed abundances.

\section{Mixing Mechanisms}

The mechanisms proposed for explaining the Li depletion in F-type stars can
be divided into two different groups: those which invoke nuclear burning of
the Li atoms and those in which this is not necessary. 

From the latter group, superficial mass loss (Schramm et al. 1990) has been
shown not to be a reliable mechanism (e.g. Swenson \& Faulkner 1992).
Microscopic diffusion (Michaud 1986; Richer \& Michaud 1993) operates where
the material sinks below the surface convection zone because the internal
radiation pressure is not capable of supporting the weight of the nuclide
concerned. Michaud (1988) suggested that if the radiative acceleration is
insufficient to support Li in the outer layers of an F-type star, the same
could then hold for nitrogen and oxygen. However, not only do the oxygen
abundances shown in Figure 2 not show any sign of depletion (either in
F-type stars of open clusters; Garc\'\i a L\'opez et al. 1993), but neither
do the radiative accelerations computed by Gonz\'alez et al. (1995) reproduce
the observed abundances for these elements. Furthermore, recent work (e.g.
Balachandran 1995) provides complementary information which suggests that the
Li depletion in old clusters is not related to microscopic diffusion.
\index{Superficial Mass Loss}
\index{Microscopic Diffusion} 

Rotation plays an important role in late-type stars of open clusters, where
high rotational velocities seem to inhibit the Li depletion (e.g. Garc\'\i a
L\'opez et al. 1994; Barrado y Navascu\'es \& Stauffer 1996). Mixing
mechanisms directly related to rotation, such as meridional circulation
(Charbonneau \& Michaud 1988) or rotationally induced turbulent mixing
(Vauclair 1988; Charbonnel et al. 1992), have been proposed within the former
group. While meridional circulation faces several problems in explaining the
Li abundances in F-type stars (e.g. Balachandran 1990), rotationally induced
mixing (after adding an important amount of pre-MS Li depletion) would
marginally reproduce the Li and Be abundances observed in the Hyades
(Garc\'\i a L\'opez et al. 1995b). 
\index{Meridional Circulation}
\index{Rotationally Induced Mixing}

Pinsonneault et al. (1989, 1990) and Deliyannis \& Pinsonneault (1992) computed
rotating evolutionary models including angular momentum loss, in which mixing
of material is associated with the transport of angular momentum in the
interior of the star. Although the models of Pinsonneault et al. can reproduce
the light-element depletion trends observed in Figure 2, they can explain the
observed Li and Be abundances in the Hyades (where the age is fixed) only if
the initial angular momenta of the late-type stars were progressively larger
for decreasing stellar mass (Garc\'\i a L\'opez et al. 1995b). Very recently,
however, Deliyannis et al. (1997; private communication) show that the
predictions from rotating models can follow the  Be vs. Li trend observed in an
independent sample of stars. A different constraint on the rotation-induced
turbulent diffusion used by Pinsonneault et al. (1989) comes from the fact that
this model fails to extract sufficient angular momentum from the radiative
solar interior to achieve the flat rotation profile revealed by helioseismology
(Brown et al. 1989).

Garc\'\i a L\'opez \& Spruit (1991) proposed a mixing mechanism for F-type
stars based on a weak turbulence induced by internal gravity waves, which is
able to explain the Li abundances in the Pleiades and Hyades, as well as the Be
measurements in the Hyades. The waves are generated by the fluctuating pressure
of the convective cells at the base of the convection zone. A different version
of this formalism was developed and successfully applied to the Sun
(Montalb\'an 1994) and to late-type stars of the Hyades (Montalb\'an \&
Schatzman 1996).
\index{Internal Gravity Waves}

However, the published mechanism for F-type stars, which was developed using
several simplifying assumptions and aimed at testing the importance of the
waves in inducing mixing in these stars, does not predict a large Be depletion
for stars of about 1 Gyr (such as those observed in Figure 2). This is now
under revision by applying a more refined convection treatment. Furthermore,
the mechanism was linked to the stellar mass and age providing abundances which
depend only on these parameters, in contradiction with the observations.
However, a high rotational velocity could change the temperature distribution
in the stellar interior (Mart\'\i n \& Claret 1996), or it could also block the
mechanical energy transfer and/or the generation of internal waves at the base
of the convection zone (Spruit 1987; Schatzman 1993), thus providing ways in
which the mixing induced by the waves is linked to the rotational velocities,
and not only to the stellar structure.

Very recently, Zahn et al. (1997) and Kumar \& Quataert (1997) have found
that internal gravity waves are also able to transport angular momentum in the
stellar interior, operating on a time scale of 10$^7$ yr for the solar case,
and being consistent with the rotation profile provided by helioseismology.

\section{Conclusions}

A similar large scatter ($>2$ dex) have been found for Li and Be abundances
in the sample of old F-type stars studied. There is a similar distribution of
light element abundances when plotted against $T_{\rm eff}$ and mass,
indicating that the depletion mechanism is not related only to the stellar
structure.

Oxygen abundances are within $\pm 0.5$ dex of the solar value (except for the
most massive star showing [O/H]$=0.73$), and appear to be correlated with
increasing stellar mass. The abundances are much higher than those predicted
by theoretical radiative accelerations and do not show any dramatic sign of
microscopic diffusion.

Beryllium and lithium have been simultaneously detected in eight stars with a
$\sim 1.5$ dex range of depletions. Assuming similar Li and Be initial
abundances for all of them (within a small range of [Fe/H]), their Li/Be
ratio would serve to constrain proposed mixing mechanisms.  

Since there are physical mechanisms linking the transport of angular momentum
and mixing of material in late-type stars to the presence and propagation of
internal gravity waves in the stellar interior, which satisfy observational
constraints imposed by the solar rotation profile and the light element
abundances in the Sun and in two open clusters, and ways have been suggested of
relating the production and/or efficiency of the waves to the stellar
rotational velocity, the internal gravity waves within the framework of
rotating evolutionary models appear as a very promising consistent explanation
for the Li and Be abundances observed in different open clusters and the field,
including the dispersions found at a given $T_{\rm eff}$ or mass. Mixing
directly associated with the transport of angular momentum in the stellar
interior is considered also as a possible explanation.


\end{document}